\documentstyle[12pt,epsfig]{article}
\begin{document}


\begin{center}
{\large \bf Exotic muon decays and the KARMEN anomaly} 
\end{center}
\vspace{0.5cm}

\begin{center}  
S.N.~Gninenko\footnote{E-mail address:
 Sergei.Gninenko\char 64 cern.ch} and N.V.~Krasnikov\footnote{E-mail address:
nkrasnik\char 64 vxcern.cern.ch}\\
{\it Institute for Nuclear Research of the Russian Academy of Sciences, Moscow 117312}
\end{center}

\begin{abstract}
An anomaly in  time
distribution of neutrinos from the ISIS pulsed beam stop source observed by the KARMEN collaboration is discussed.\ We show that the anomaly 
 can be interpreted as a superposition of two exponentials, both
having  time constants  consistent with the $\mu^{+}$ lifetime of 2.2 $\mu s$.\
It is assumed that they both originate from  muon decays at rest.\ 
 One of them 
describes the  
time distribution of the prompt  neutrino events, while the other describes the
 time distribution of events from delayed decays of  
slowly moving ($\beta \simeq 0.02$) particles in the KARMEN calorimeter.\ 

We propose here that these particles are produced in exotic
decays of positive muons $\mu^{+}\rightarrow e^{+} + X$,
resulting
in the second exponential time distribution  shifted by the time of flight with 
respect to 
the time distribution of neutrino events.\ This model gives an aceptable fit 
to the KARMEN data if $X$ has a mass of 103.9 MeV.\
  The possible 
decay modes  of this new massive neutral particle  are discussed.\ 
This hypothesis can be experimentally tested in the near future by studying 
the low energy part of the $e^{+}$ spectrum in the $\mu^{+}$ decays.   
\end{abstract}

\section{Introduction}

The KARMEN collaboration, which studies the interactions of neutrinos from
the stopped $\pi^{+}$ decay chain at Rutherford Appelton Laboratory, 
has reported an anomaly \cite{1} in the time distribution of
neutrino events in the KARMEN calorimeter.\ 
There is a bump of events in the distribution which was expected to be well 
described by a single exponential with a time constant equal to the muon
decay lifetime.\  

It was suggested that a new 
weakly - interacting half spin particle with a mass 33.9 MeV/c$^{2}$ from initial $\pi^{+}\rightarrow \mu^{+} + x$ decay
, which travels with velocity $\beta = v/c \simeq 1/60$ and decays in the 
KARMEN calorimeter after a mean flight path of 17.5 metres, can be a possible 
candidate for an explanation of this anomaly.\ 
Since the visible energy in the calorimeter was in the range from 11 to 35 MeV 
it was also assumed  that these new particles deposited  
only a fraction of their full energy in the calorimeter.\ The hypothetical $\pi^{+}\rightarrow \mu^{+} + x$ decay has been searched for at PSI by studying muons from
pion decay in flight \cite{2}.\ The result gives a branching ratio of less 
than $7\times 10^{-8}$ at 95$\%$ Confidence Level (C.L.) for this decay mode 
leaving only a narrow allowed region in the branching ratio if combined 
with the results of ref. \cite{3}. 

  Another possible interpretation of the anomaly proposed by the 
KARMEN collaboration is related to the 
direct production of a new particle in the proton collision with the 
neutrino target. However, the collaboration has not yet stated whether it has
devised one.

Concerning the possible explanation of the anomaly by exotic muon decays it 
was concluded 
that ``...the narrow clustering of the anomalous events
rules out contributions from $\mu^{+}$ decay,
which spreads over a much longer time interval (compared to the 
$\pi^{+}$ - life  time of  $\approx 20~ ns$) up to 10 $\mu s$ after beam-on
target.'' \cite{1}.\ 

  This conclusion is the  motivation for present letter, where we show that 
 the KARMEN anomaly could be explained by contributions from 
the exotic $\mu^{+}$ decays. 

The paper is organised as follows. 
In section 2 we analyse the time distribution obtained by the KARMEN 
collaboration and give a possible explanation of the anomaly.\ As an example we consider the $\mu^{+}\rightarrow e^{+} + X$ decay
  into a positron and a new boson  with the mass  
of 103.9  MeV/c$^{2}$.\
In sections 3,4 the possible X-boson phenomenologies  
 and experiments 
on searching for the rare decay $\mu^{+}\rightarrow e^{+} + X$ are discussed,
respectively.\ Section 5 contains concluding remarks.\

\section{Analysis of the neutrino time spectrum}

Figure 1 shows the time distribution found by KARMEN with respect to beam-on 
target events.\ Analysis of this time distribution 
 was performed in ref.\cite{1} after 
subtraction of the cosmic ray induced background, which was found to be 
constant and equal to 60 events per 0.5 $\mu s$.\footnote{ The experimental
points  were reproduced from the original
KARMEN publication \cite{1}. The error bars were calculated  taking into 
account the errors due to subtraction of the background.} 

 This spectrum was fitted in ref. \cite{1} by 
a superposition of Gaussian  and exponential signals from muon decay 
neutrinos having a 2.2 $\mu s$ decay constant.\ 
The fit 
procedure performed with the minimization routines of MINUIT \cite{4} 
results in $\chi^{2}$ of 9.7 for 14 degrees of freedom (DF)
corresponding  to  79$\%$ C.L. for this hypothesis.\
Gaussian, which is  centred at $(3.6\pm 0.25) \mu s$ with a full width 
at half maximum(FWHM) of $(1.3\pm 0.5) \mu s$  consisted
of 83$\pm$28 events  associated with the signals 
from decay in flight of the proposed particle in the fiducial volume of the 
KARMEN calorimeter.\

We tried to reproduce these results by using the same MINUIT minimizing routines.\  Our results  and results of the
 KARMEN fit were found to be in a good agreement: our fit procedure yielded
 $\chi^{2}$ of 9.8 for 14 degrees of freedom (DF)
corresponding  to 77$\%$ C.L. for this hypothesis.\
Gaussian was found to be  centred at $(3.7\pm 0.26) \mu s$ with a FWHM
 of $(1.3\pm 0.5) \mu s$ and with 
85$\pm$32 events.\ The result of our fit procedure is shown in Figure 1.\

Having this good agreement in mind,  
we  then performed an exponential fit of the KARMEN time distribution 
assuming that this spectrum is described by a sum of 
 two exponentials: the first exponential describes the time 
distribution in the interval 1 - 3.3 $\mu s$ , and  the second one 
describes the time distribution in the interval 3.3 - 10 $\mu s$.\ 
Figure 2 shows the result of the fit (solid lines).\
The MINUIT
minimizing routines yield a $\chi ^{2}$ of 9.7 for 15 DF  corresponding 
to 84$\%$ C.L., and time constants of 
$(2.29\pm 0.34) \mu s$ and $(2.1\pm 0.6) \mu s$ for the first and 
the second exponentials respectively.\ 
Note that there is no significant difference in 
Confidence Levels of our and KARMEN's hypotheses.

The time width of the transition region between two exponentials is rather
 small and corresponds to 
$\simeq 0.5 \mu s$.\ We try to fit this region more precisely  using 
a Gaussian shape of the transition curve instead of a step-function.\ 
However, no significant difference between two results has been found. 
 
A similar description of the time distribution was obtained by superimposing
an exponential describing the distribution of events in the time interval 
3.3 - 10 $\mu s$ upon an exponential 
signal having a 2.2 $\mu s$ decay constant for the full time interval.\  
Figure 3
shows the corresponding time distribution of events after subtraction of the
 exponential part with 2.2 $\mu s$. 
A direct exponential fit of 
the time distribution of events in this region  
 gives a time 
constant of 2.26 $\mu s$  close to the value of muon decay 
constant; however the error is rather large because of poor statistics.

 Since the time constants were found to be equal within the error limits
to the muon decay constant we may assume  that  all events in the KARMEN
time distribution  originate  from the muon decays.\ 

   Thus, our interpretation of the KARMEN anomaly is the following.\ 
 Positive muons from $\pi^{+}$ decays
stop in the target ( muon range in the target is $\simeq$ 1 cm) and decay 
giving a flux of neutrinos which decreases in time exponentially with the 
muon decay constant.\ 
The first 
exponential shown in the time spectrum in Figure 2 is related to the interactions of this flux
 of neutrinos in the KARMEN calorimeter.\
An excess of  events  after $\simeq $ 3.6 $\mu s$ shown in Figure 3 
is related to the decays of slow moving particles  produced in  
muon decays.\
Therefore, the flux of these particles and fraction of their decays in the 
fiducial volume of the KARMEN calorimeter also decrease in time exponentially 
with the muon decay constant, resulting in the second time distribution but
shifted by the time of flight with respect to the time distribution of 
neutrino events.\ Since the time width of the transition region 
between two exponentials  is rather narrow (see Figure 2), we suggest that 
these new particles originate from  the two-body decays of positive muons, 
namely $\mu \rightarrow e^{+} + X$.\ We note that the RMS of fluctuations of 
time needed for $X$ bosons to pass from the neutrino target to  
the calorimeter and  decay inside its volume was estimated to be of the 
order of 0.2 $\mu s$.\ The effect of the convolution of these fluctuations and
the exponential from muon decay on the exponential shape of this curve and its
decay constant  was found to be small.\  

The mass of the $X$ boson calculated for $\beta_{X}=1/60$ is 
found to be (103.9$\pm$0.1) MeV/c$^{2}$, where the error corresponds to
 the spread of $X$'s velocity estimated from the time width of the transition region between two exponentials.\  
The excess of events after subtraction 
of the contribution from the neutrino was found to be  
$\Delta N = 108 \pm 41$ events.\ 

The limit on branching ratio 
$Br = \Gamma(\mu \rightarrow e^{+} + X)/\Gamma(\mu \rightarrow all)$ as a function of the $X$-boson lifetime 
can be recalculated from the KARMEN result for $\pi^{+} \rightarrow \mu^{+} +
x$ decay mode taking into account the difference in the number of excess events 
found after the fit procedure.\ The result is shown in Figure 4.

  
\section{ Possible X-boson phenomenologies}

Suppose that X-boson is a neutral scalar particle. The Lagrangian 
describing the  $\mu \rightarrow e + X$ decay has the form
\begin{equation}
L = X[h_{\mu e L}\bar{\mu}_Le_R + h_{\mu e R}\bar{\mu}_Re_L + h.c.]
\end{equation}
The $\mu \rightarrow e + X$ decay width is given by the formula 
\begin{equation}
\Gamma(\mu \rightarrow e + X) = \frac{ m_{\mu}}{32\pi}(h^2_{\mu e L} +
h^2_{\mu e R})(1 - \frac{m^2_X -m^2_e}{m^2_{\mu}})\cdot 
[(1 + \frac{m^2_e - m^2_{X}}{m^2_{\mu}})^2 - 
4 \frac{m^2_e}{m^2_{\mu}}]^{\frac{1}{2}}
\end{equation}
For a non-relativistic X-boson with a speed $v_X = \frac{1}{60}c$ we 
find that its mass is $m_{X} = 103.9$ MeV and the $\mu \rightarrow e + X$ 
branching ratio is 
\begin{equation}
Br(\mu \rightarrow eX) \approx 0.4 \cdot
10^{13}[h^2_{\mu e L} + h^2_{\mu e R}]
\end{equation}
For the interaction (1) the X-boson decays via the virtual $\mu$ meson 
$X \rightarrow (\mu^{*} \rightarrow e \nu \bar{\nu}) + e$ and its 
decay width is 
\begin{equation}
\Gamma(X \rightarrow e^{+}e^{-} \nu \bar{\nu}) \sim 
O((\pi^2 \Gamma(\mu \rightarrow e \nu \bar{\nu})
(h^2_{\mu e L} + h^2_{\mu e R}))
\end{equation}

As it follows from the observed number of the anomalous events due to the 
X-boson decay in the KARMEN experiment, the ratio 
\begin{equation}
K \equiv  Br(\mu \rightarrow e + X)\Gamma(X \rightarrow vis)
\end{equation}
is   
\begin{equation}
K \sim O(3 \cdot 10^{-11}s^{-1})
\end{equation}
From the formulae (2 -6) we find that $Br(\mu \rightarrow e + X) \sim 
O(10^{-2})$ and $\tau(X \rightarrow \mu e \nu \bar{\nu}) \sim O(10^9) s$, 
that is probably at the level of the contradiction with the  experimental 
bound \cite{6} $Br(\mu \rightarrow e + X) \leq 3 \cdot 10^{-4}$ which is valid 
however for $m_{X} \geq 103.5$ MeV (see Section 4).\ It is possible to 
invent a model where the 
$Br(\mu \rightarrow e + X)$ is less than $O(10^{-4})$.\ Let us introduce 
the hypothetical heavy charged lepton L and consider the decay of the 
X-boson through the virtual charged L-lepton
\begin{equation}
X \rightarrow (L^{*} \rightarrow e \nu \bar{\nu}) + e
\end{equation} 
For the L-lepton mass $m_{L} = 100$ GeV, the coupling constant of L-lepton 
with 
electron and X-boson $h_{LeX}=0.1$, the L-lepton mixing with electron 
$\epsilon^{2}_{eL} = 10^{-2}$, we find that $Br(\mu \rightarrow  e X) = 
O(10^{-5})$.\ So, if our interpretation of the KARMEN anomaly is correct we 
expect a rather large branching ratio  
$Br(\mu \rightarrow e + X) \geq O(10^{-5})$.\ Therefore, 
it would be very interesting to perform an experiment 
 on the search for $\mu \rightarrow e + X$ decay with improved sensitivity.\
 These scenarios predict the X-boson 
lifetime $\tau(X) \geq O(10^6)$ s that probably leads to problems with
astrophysical and cosmological constraints. Moreover, these scenarios 
predict an anomalously high X-boson decay energy $<E_{vis}> \sim 50$ MeV 
compared to the observed visible energy $E_{vis} \sim (11-35)$ MeV. 

To avoid these problems, let us consider the model with two additional neutral 
scalar bosons $\phi_{1}$ and $\phi_{2}$ and suppose that the $\phi_{1}$-boson 
is lighter than the X-boson and the $\phi_{2}$-boson is heavier than the 
X-boson. The Lagrangian describing the X-boson decay has the form 
\begin{equation}
L = \lambda X \phi_{1} \phi_{2} + h_e \phi_{2}\bar{e} e + 
h_{\nu} \phi_{1} \bar{\nu} \nu .
\end{equation}
The X-boson decays via the reaction 
\begin{equation}
X \rightarrow \phi_{1} + (\phi^{*}_{2} \rightarrow e^{+}e^{-})
\end{equation}
through the virtual $\phi_{2}$-boson and the neutral $\phi_{1}$-boson 
decays into neutrino antineutrino pair. For such a scenario it is possible 
to have $<T_{vis}> \sim 30 $ MeV and $\tau(X) \leq 10^{-2}$ s. For 
instance, for $m_{X} = 103.9$ MeV, $m_{\phi_{1}} =73$ MeV, 
$m_{\phi_{2}} = 1$ GeV, $h_{e} = 10^{-3}$, $\lambda = 10^{-2}m_{\mu}$ 
and $h_{\bar{\nu}} = 10^{-6}$ we find that the average visible energy 
in the X-decay is $<T_{vis}> \approx 26$ MeV and 
$\tau(X) \sim 3 \cdot 10^{-3}$ s, 
$\tau(\phi_{1}) \sim 2 \cdot 10^{-10}$ s, $Br(\mu \rightarrow e + X) 
\approx 10^{-13}$, $\tau(\phi_{2}) \approx 3 \cdot 10^{-17}$ s. 
So in this  rather artificial model all particles are not long-lived and 
it is possible to have $<T_{vis}>$  in agreement with KARMEN data. 

Note also that it is possible to have a model where the $\phi_{2}$-boson 
is the Higgs boson. For instance, for the Higgs boson mass $m_h = 100$ GeV 
and $\lambda = m_{\mu}$ we find that $\tau(X \rightarrow \phi_{1} e^+ e^-) 
\sim 0.7 \cdot 10^{7}$ s and $Br(\mu \rightarrow eX) \sim 2 \cdot 10^{-4}$. 
In this case to have $\tau(X) \leq 10^{-2}$ s we 
have to postulate an additional interaction of the $X$-boson with neutrinos.

\section{Experimental search for the  $\mu^{+}\rightarrow e^{+}+X$ decay}

 Since  $X$'s are weakly-interacting and relatively  long lived particles, 
their flux  would penetrate any type of calorimeter 
without significant attenuation  and cannot be  
observed effectively in the detector via its decays.\ A more effective 
experimental signature for $\mu^{+}\rightarrow e^{+}+X$ decays  would be a
peak from mono-energetic positrons in the low  energy part of the 
positron spectrum corresponding to the principal mode of $\mu ^{+}$ decays.\

Direct experimental searches for a peak from 
$\mu^{+}\rightarrow  e^{+}+X$ decays were   performed in a
few experiments \cite{5}.\ 
For the lowest positron energy region from 
1.6 to 6.8 MeV, corresponding to the mass region of 103.5 - 98.3 MeV/c$^{2}$, the branching ratio limit of $\Gamma(\mu^{+}\rightarrow e^{+}+X)/\Gamma(\mu^{+}\rightarrow all) \leq 3\times 10^{-4}$
was obtained in ref. \cite{6} from the analysis of published data on muon 
decays in the hydrogen bubble chamber obtained by Derenzo \cite{7}.\ The result is limited by a poor ( $\simeq$ 30-40$\%$) momentum resolution in this region.

It would be interesting to search for a peak in the low energy part of the 
positron spectrum by using, for example, a high resolution Ge-detector (
 NaI detector) with a
typical  energy resolution of the order of KeV ($\simeq$ 100 KeV).\ The 
 low energy
muons could be stopped in the active volume of the detector.\ In this case 
the low energy spectrum  
 of the positrons  could  be measured without significant 
distortions.\  

 The estimate of the rate from the principal muon decays  shows that, using 
Ge-detector, one 
could expect a sensitivity to the $\mu^{+}\rightarrow e^{+}+X$ decay
branching ratio of better than $10^{-6}$.\ Here it is assumed that the intensity 
of stops is of the order of $10^{4}\mu^{+}$/s and that the  
exposure to the muon beam is  $\simeq$ 1 month.\

\section {Conclusion}

The anomalous time distribution of the events from the KARMEN detector
can be acceptably  described 
by a superposition of the 
time distribution of neutrino signals following the expected 2.2 $\mu s$
muon decay constant  and a weaker exponential signal from delayed decays of   
the new massive neutral particles also  originated from the muon decay
at rest and shifted in time due to their time of flight.\ 

 It is proposed  that a new boson $X$ with the mass of  103.9  MeV/c$^{2}$
which might be produced in the
decays of positive muons $\mu^{+}\rightarrow e^{+} + X$
could be a possible candidate for an explanation of the anomaly.\ 
  The visible energy from the possible decay modes of  $X$'s can be 
within the allowed region from 11 to 35 MeV/c$^{2}$.\ 
 This hypothesis can be experimentally tested in 
the near future at a level of sensitivity of better than 10$^{-6}$.\ 

It should be noted that the original interpretation of the KARMEN anomaly 
as a new neutral fermion $x$ produced in $\pi^{+} \rightarrow \mu^{+} x$ 
decay with a mass $m_{x} = 33.9$ MeV has been discussed in ref. \cite{3}. 
It has been shown that a mainly sterile neutrino scenario is compatible 
with all laboratory constraints, although there are problems with 
astrophysical and cosmological constraints. Our interpretation with the scalar 
$X$-boson produced in the $\mu^{+} \rightarrow e^{+}X $ decay is in the spirit 
of the familon scenario \cite{8}.

\vspace{2.0cm}

{\Large \bf Acknowledgements}

\vspace{0.5cm}

 We thank Prof. V.Matveev for useful discussions and  remarks, 
Prof. F.Wilczek for useful comment and  
A.Toropin for help in  calculations.\    
One of the authors (S.N.G.) would like to thank his colleagues from the 
NOMAD collaboration at CERN for fruitful discussions.\

\newpage

 \begin{figure}
   \mbox{\hspace{-1.5cm}\epsfig{file=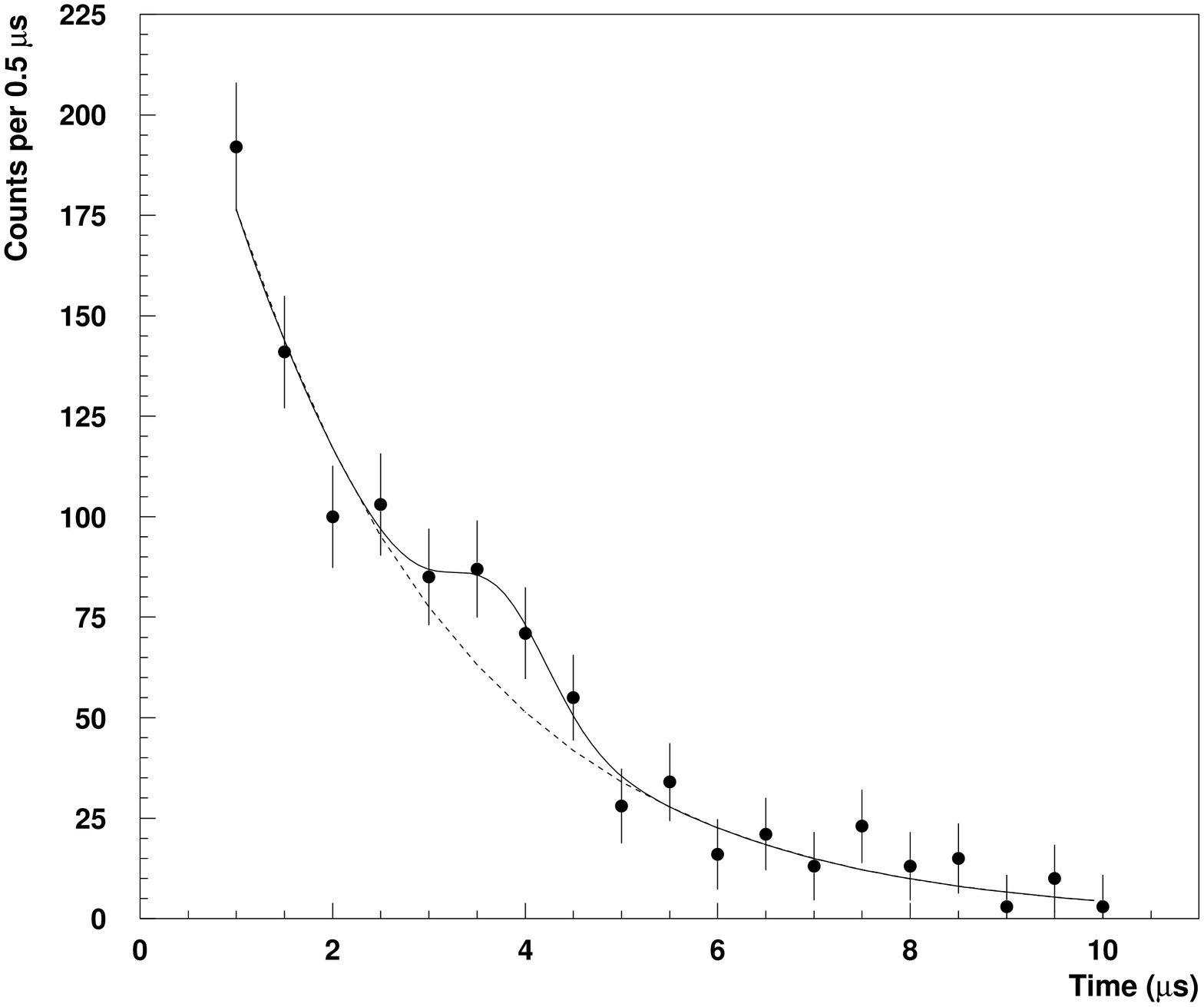,height=150mm}}
    \centering
  \caption{\em Time distribution of events in the 
KARMEN calorimeter after the subtraction of the cosmic background.$^{3}$ The data 
 are fitted to an exponential with the 2.2 $\mu s$ decay constant on which
is superimposed a Gaussian signal centred at 3.7 $\mu s$.\ The fit procedure 
results in $\chi^{2}$ of 9.8 for 14 degrees of freedom.}
  \label{mu1}
\end{figure}

\newpage

\begin{figure}[hbt]
\begin{center}
   \mbox{\hspace{-1.5cm}\epsfig{file=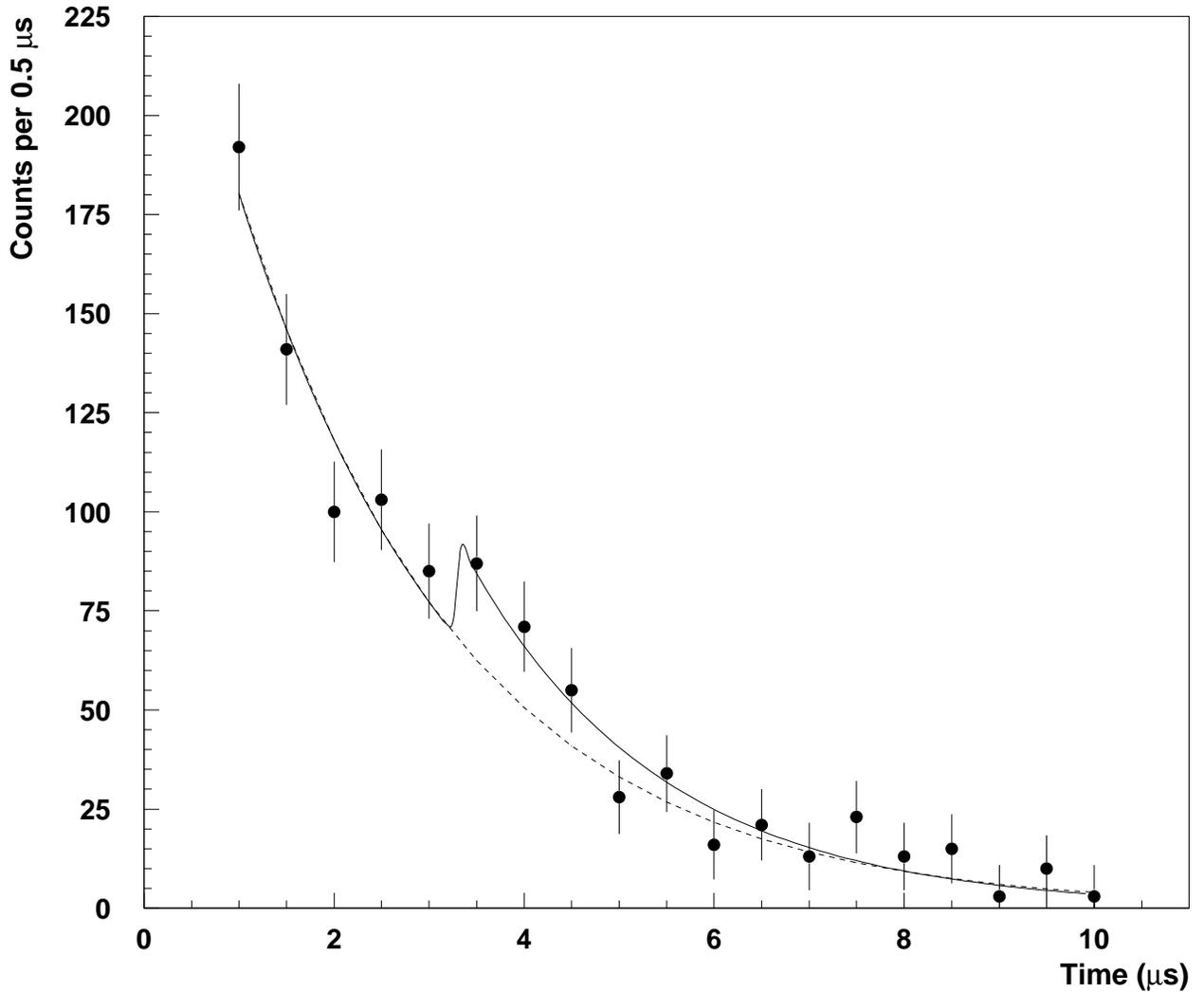,height=150mm}}
  \caption{\em Time distribution of events in the 
KARMEN calorimeter after the subtraction of the cosmic background.$^{3}$ \ The solid 
curves are a fit
to the points by a sum of two exponentials.\ The first exponential describes
 the time distribution 
in the region from  1.0 to 3.3  $\mu s$ and the other in the region from
 3.3 to 10  $\mu s$ with time constants of $(2.29\pm 0.34) \mu s$ and $(2.1\pm 0.6) \mu s$,
respectively.\ The broken line corresponds to the extrapolation of the first
exponential.\  The fit procedure 
results in $\chi^{2}$ of 9.7 for 15 degrees of freedom.}     
\end{center}
  \label{figure 19:}
\end{figure}

\newpage

 \begin{figure}
   \mbox{\hspace{-1.5cm}\epsfig{file=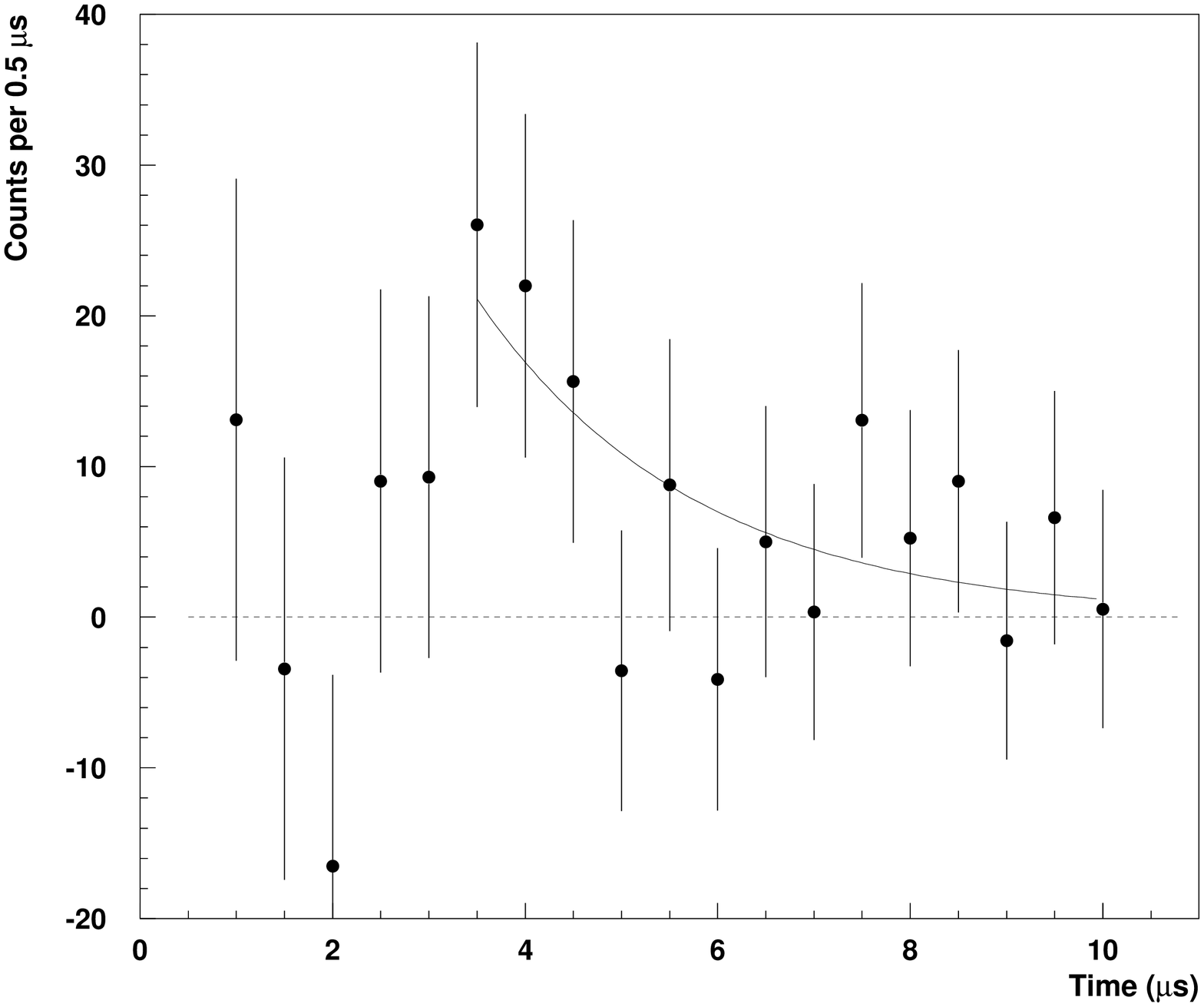,height=150mm}}
    \centering
  \caption{\em Time distribution of excess events left after subtraction 
of the exponential component with 2.2 $\mu s$ decay constant. The solid line 
is the exponential fit to the points in the interval 3.3 - 10 $\mu s$ which 
gives a time constant of 2.26$\mu s$.}
  \label{mu1}
\end{figure}

\newpage

 \begin{figure}
   \mbox{\hspace{-1.5cm}\epsfig{file=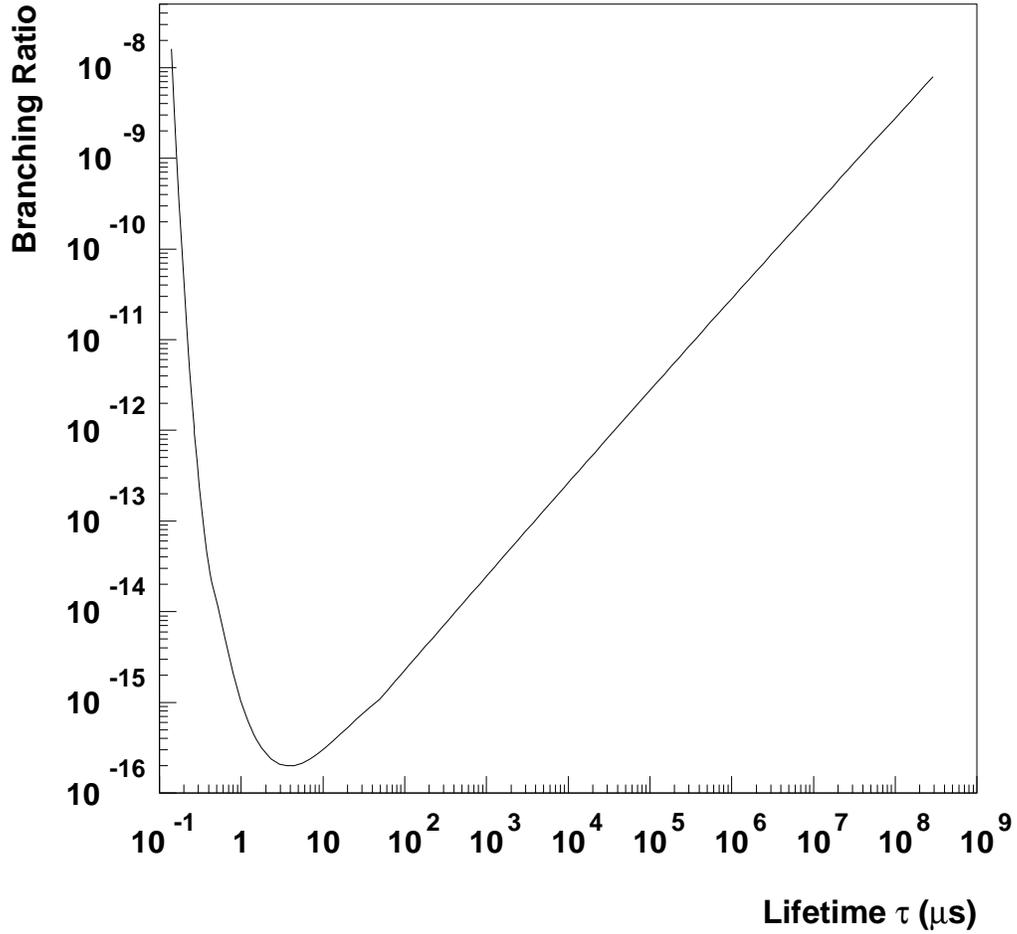,height=150mm}}
    \centering
  \caption{\em The branching ratio $\Gamma(\mu^{+}\rightarrow e^{+}+X)/\Gamma(\mu^{+}\rightarrow all)$ as a function of the lifetime of the $X$-boson calculated
 from the number of observed events ( $\Delta N = 91$ events) using the analogous
KARMEN plot for a decay mode of $\pi^{+} \rightarrow \mu^{+} + x$.}
  \label{mu1}
\end{figure}

\end{document}